\documentclass[twocolumn,10pt]{asme2ej}

\usepackage{epsfig} 

\title{Navigating Surveillance Capitalism: A Critical Analysis through philosophical perspectives in Computer Ethics}

\author{Angelica Sofia Valeriani
    \affiliation{
    Ethics of Information Technology \\
	Politecnico di Milano, Milan, Italy\\
    }	
}

\usepackage{url}
\usepackage[hidelinks]{hyperref}
\usepackage{amsmath}
\usepackage{amssymb}

\begin{document}

\maketitle    

\begin{abstract}
{\it
Surveillance capitalism is a concept that describes the practice of collecting and analyzing massive amounts of user data for the purpose of targeted advertising and other forms of monetization. The phenomenon has become increasingly prevalent in recent years, with tech companies like Google and Facebook using users' personal information to deliver personalized content and advertisements. Another example of surveillance capitalism is the use of military technology to collect and analyze data for national security purposes. In this context, surveillance capitalism involves the use of technologies like facial recognition and social media monitoring to gather information on individuals and groups deemed to be potential threats to national security. This information is then used to inform military operations and decision-making. This paper wants to analyze in a critical way the phenomenon of surveillance capitalism, proposed under two different ethical framework perspectives. Utilitarianism, a consequentialist ethical theory that judges actions based on their ability to bring about the greatest amount of happiness or pleasure for the greatest number of people, and Kantian deontology, a non-consequentialist ethical theory that emphasizes the importance of individual autonomy, freedom, and dignity. On one side, the utilitarian framework enlightens how Information Technology (IT) and the features provided offer, at first sight, all the positive perceptions to the majority of people, happiness, entertainment, and pleasure. On the other side, the Kantian deontology framework mostly focuses on the aspect of freedom and free will of the individual. This topic is  particularly related to the concession of permissions to access data in change of services and the degree of influence that manipulation performed by surveillance capitalism can generate.}
\end{abstract}



\section{Introduction}
To increase consumerism, the business of major companies and the power of countries, social networks and political parties are performing a form of inner and intimate manipulation that is located in the idea of Surveillance Capitalism. In this paper, I will try to show this new phenomenon, which can be realized utilizing targeted advertising and military technology, under two different ethical frameworks, the utilitarianist and the Kantian perspective. The scope is to enlighten justifications and implications according to these different frameworks. In particular, I will start by defining the concept of Surveillance Capitalism and the way it is perceived nowadays. I will analyze in depth some of the most common real-world situations in which this phenomenon can be seen in action, i.e. targeted advertisement and coercive manipulation in both totalitarian regime and democratic countries. I will then analyze these aspects in relation to the ethical frameworks proposed, from one side “the end justifies the means” and from the other the duty framework as the core, introducing other philosophical works as a support of my analysis. 

The paper is organized in the following way. Section \ref{surveillance} will describe Surveillance Capitalism, defining its characteristics and important features. Section \ref{ads} and \ref{war} will be devoted to the analysis of the realization of Surveillance Capitalism in the form of targeted advertisement and manipulation of people in politics and military fields. In particular, these real scenarios of analysis were chosen because they are very representative of the main needs in the nowadays society. From one side the target advertisement represents an incentive for consumerism and the possibility of easily realizing every desire (e.g. the easy discovery of new places, occasions, and assets). From the other side, the social and national security, which is connected to both politics and military war is a topic to which people are very sensitive nowadays, because of the incredible ease with which some crimes can be performed today. Section \ref{util} recalls these phenomena, in particular the one of target advertisement, which is more connected to pleasure and amusement, under a utilitarian perspective, after having described the key points of the framework. Section \ref{duty} will have the same approach as section \ref{util}, applied to the duty framework proposed by Kant and focusing more on the phenomena related to the political sphere, as it is more representative of the concepts of responsibility and duty towards people and the country. Section \ref{comp} will complete the analysis, deepening the study of both phenomena under the two ethical frameworks. It will also enlighten the main difference between the two frameworks and their values, while conclusions are drawn in section \ref{end}.

\section{The new Empire of Surveillance Capitalism}
\label{surveillance}
The wide spread of the Internet brought to the current society in which the major means of generating wealth on the Internet and through proprietary platforms (as apps) is the surveillance of the population. This phenomenon allows to increase exponentially gains from the digitalized companies that have the monopoly of society. Digitalization of surveillance has radically changed the nature of advertising. Now, the implication of the system is the absence of effective privacy. Revelations of Edward Snowden on NSA’s Prism of 2013 are an example of the pattern of a tight interweaving of the military with giant computer-Internet corporations. There are many examples of partly cooperative, partly legally coerced sharing of data that can be found in the case of Microsoft, Google, Yahoo, Facebook, and others. These companies turned over the data from tens of thousands of their accounts on individuals every six months both to the NSA and other intelligence agencies, with a fast rise in the number of accounts turned over to the secret government \cite{Surveillance1}. In practice, according to revelations, NSA gained access to data from mobile phones emanating from hundreds of millions of Americans as well as populations abroad—operating thorough Boundless Informant, Prism, and other secret projects. The final goal was to capitalize on new military technology and create larger global Internet monopolies while expanding the military-digital interchange system. In the context of capitalization, an example of the trend that well represents the centralized structure of monopoly-finance capital in the age of digital surveillance is given by the practice of “securitization” increasingly standing simultaneously for a world dominated by the elements identified in \cite{Surveillance1, Surveillance3}. In detail:
\begin{enumerate}
  \item Financial derivatives trading
  \item A network of public and private surveillance
  \item The militarization of security-control systems
  \item The removal of judicial processes from effective civilian control
\end{enumerate}
\begin{figure}[t]
\centering
\includegraphics[trim=0mm 0mm 0mm 0mm,clip,width=\linewidth]{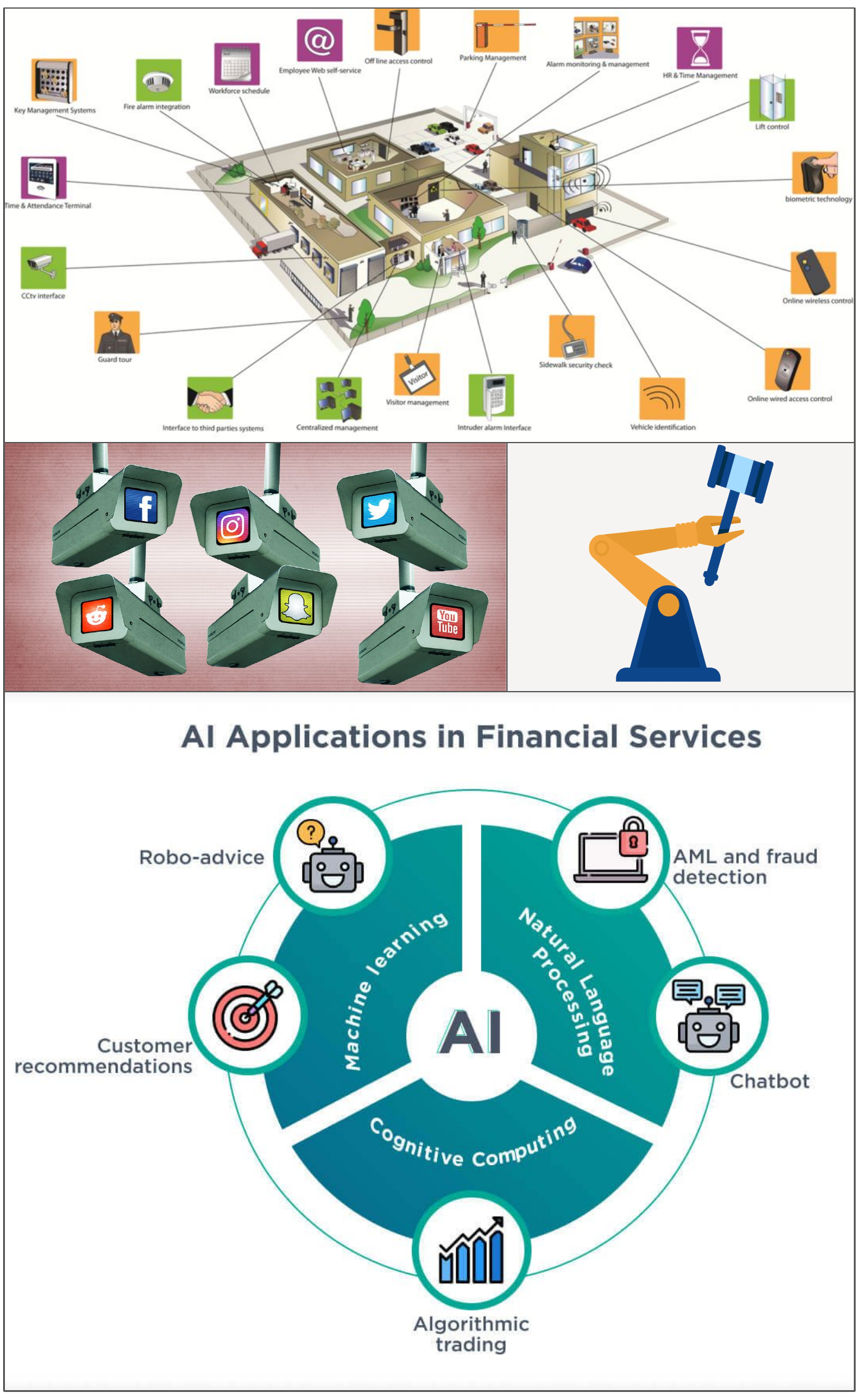}
\caption{Categorization of main Surveillance fields. Security-control systems, public and private surveillance, AI applications in legal courts and financial services.}
\label{framework}
\end{figure}
Surveillance can be simply seen as a collection of techniques that together potentially constitute subjects to regulation (first interpretation), as well as attention purposeful, systematic, and oriented to social control (second interpretation). While totalitarian regimes may embrace the second definition, seeing Surveillance as an instrument to strengthen their control over people, capitalist and democratic systems may value Surveillance more for reasons connected to its rationality, potentialities in terms of welfare, and good sense. On the whole, it must be noticed anyway that Surveillance must be contextualized also in its economic implications and this is more than a way of social control \cite{Surveillance2}. One of the greatest issues that are connected to Surveillance Capitalism, as also shown in the core phenomenons of the paper, i.e. target advertising and coercive manipulation, is the asymmetry in the distribution of power that is generated by the retrieval of data of people unaware of this process (known and discussed following some whistle-blowers revelations). The weight of this power is concentrated in the hands of the actors that have access and can manage the information that is retrieved \cite{Surveillance3}.

\section{Target Advertisement}
\label{ads}
The way of performing advertising today has changed a lot over time. New technological instruments, such as social networks in particular, have led to a disclosure of personal data, whether sometimes intentionally performed by users, and most times unintentional, that brought a completely new way of making advertising. Targeted advertising has become a widespread marketing strategy, utilizing data-driven methods to identify potential customers with specific interests, behaviors, and characteristics. The aim is to deliver personalized advertising messages to the right audience, at the right time and place. It is important to notice that in this scenario companies and agencies are only interested in maximizing their business as much as possible. Their primary and only goal is to refine their marketing strategies and increase consumption from users. Typically, when the individual sets up their own profile on a social network, the provided information is collected, stored, and then used by promoters for easier reach to more specialized groups of the targeted audience. Consumer behavior and habits are so interesting for companies’ businesses that they are still active in paying a huge amount of money for applications that generate databases to store such personal data, despite some rules that have been defined in order to guarantee the avoidance of exploiting data \cite{ads1}.

According to various research works and general common sense, it can be noticed how the targeted advertisement can be perceived in a different way according to the specific user. Target advertising can provide a better user experience by delivering more relevant and useful information to customers. For example, if someone is looking for a specific product or service, they are more likely to appreciate advertisements that are related to their search query. Moreover, targeted advertising can also lead to the discovery of new products and services that customers may not have been aware of otherwise, potentially resulting in positive outcomes for both customers and businesses. Hence, targeted advertising mainly allows companies to allocate their resources more efficiently by focusing on the most promising customers rather than wasting resources on those who are less likely to be interested. On the other hand, by analyzing data on people's interests, preferences, and behaviors, companies can influence customer decisions and actions, potentially leading to exploitation or manipulation. This can be especially concerning in situations where vulnerable individuals or groups are targeted, such as children or those with addiction or mental health issues. Indeed, while some people can find amusement, pleasure, and appeal in finding ads that are more connected to their interests, other people can perceive this intrusion as both an annoying phenomenon and a violation of personal privacy \cite{ads2, ads4}. 

Nowadays, research works have great expectations and focus a lot on the possibility to anonymize complete data that are taken to do business, but, at least today, this is often not possible \cite{ads3}. In target advertisement, as it is performed nowadays, different criteria that require data taken from the user are necessary. This point does not regard only the explicit ratings given to a product, but also the number of times a website is accessed, whether an advertisement is opened or not, the clicking on a link, the connections between different items of interest, the specific times of the day in which the user typically access and so on. The list of significant factors in advertising campaigns is very long and it finds its core point and strength in social media \cite{ads4}. One of the primary concerns associated with targeted advertising is privacy. The collection and use of personal data, such as browsing history, search queries, and location, can lead to privacy violations and potentially harmful consequences. For example, data breaches can result in identity theft, financial fraud, and other cybercrimes. Moreover, targeted advertising can also lead to the creation of echo chambers, where people only see information that reinforces their existing beliefs and biases, potentially leading to social and political polarization.
\section{Military technology and Politics}
\label{war}
The type of targeting described in section \ref{ads} does not have only an economic surveillance effect, but also a great impact on the phenomenon of psychological surveillance, in the sense that technological means are used to enforce manipulation and access to personal and critical information. In this section, the focus is on the political and social order scenario. A main critical aspect, under the perspective of freedom and psychological imposition, is the fact that people are obliged to allow complete access to their data in order to gain access to necessary services. Apparently, they have the choice to deny permissions, but, in practice, they have not, because the majority of services require consent to complete access to data and this implies that denying permissions forbids the user to adopt both a specific necessary service and all the similar services that could substitute it \cite{duty1, zub}. The lack of freedom is not shown only in such a mechanism, but also in other cases, related to the political and social context. 

A practical example is related to video surveillance modules based on appearance-based person detection. Here, data are manipulated and processed to get features like skin detection. The use of these instruments can have very negative effects in contexts like totalitarian regimes, as they increase the oppressing power of the government and its ferocious control grip over both the mass and the opponents to the regime \cite{war2}. Surveillance technology has transformed the way military organizations operate, providing new capabilities for intelligence gathering, reconnaissance, and situational awareness. Drones, satellites, and other remote sensing technologies have enabled military organizations to monitor large areas and gather intelligence without putting troops in harm's way. This has been particularly valuable in areas with hostile terrain or where it is difficult to gather intelligence using traditional methods. However, the use of surveillance technology in the military also raises concerns about privacy and ethical implications. The use of unmanned drones, for example, has been controversial due to the potential for civilian casualties and the lack of accountability for remote operators. Moreover, the use of surveillance technology in conflict zones can infringe on individuals' privacy rights and lead to human rights violations.

As another example, considering now democratic countries, the improper use of Surveillance Capitalism instruments in the government field can be found in the field of elections and votes. Surveillance technology has also become an increasingly common tool in the political field, where it is used to monitor individuals and groups for a variety of purposes. Governments can use surveillance technology to monitor and track dissenting voices, both domestically and abroad, potentially leading to violations of human rights and civil liberties. This can have a chilling effect on freedom of speech and the ability of individuals to express their opinions without fear of retribution. Moreover, the use of surveillance technology in political campaigns has raised concerns about the manipulation of public opinion. Data collected from social media, for example, can be used to create detailed profiles of individuals and target them with personalized messages, potentially leading to manipulation and exploitation. This can undermine the democratic process by creating an uneven playing field and giving certain candidates or parties an unfair advantage. In fact, audiences in the political sphere can be influenced and forced into decisions that can lead to contradictions with people’s actual ideas. Obviously, persuasion and manipulation have always been part of the political and marketing worlds, but in this new Surveillance Capitalism, the effectiveness of this phenomenon is amplified. All choices can be manipulated by the external environment, as a psychological mechanism; the decision-making process is a complex procedure in which Surveillance Capitalism gains a great influence. In this scenario, personalized and targeted messages, but also fake news produced ad hoc by opponents, can undermine voter autonomy and change the course of history \cite{war3, war4}. 

On the whole, it must be noticed that the first and primary reason for which many countries invested so much in this intrusion into private life was to face the threat of cyber war that is directed at both entire military and financial systems. The most representative example is the one connected to terrorist attacks and the strong intention to prevent them and stem them. This goal is only reachable by increasing security control, so from this point of view, an intrusion into the ideals of people is very effective, even if not morally correct, to sooner discover eventual threats \cite{war4}. The same reasoning can be applied to the whole financial system, as attacks directed to this field are able to seriously persuade the majority of people, for example, losing all their assets and leading to a deep and spread crisis. Surveillance technology has had a significant impact on military and political fields, providing new capabilities for intelligence gathering and decision-making. However, it also raises concerns about privacy, ethics, and the potential for abuse. It is important for policymakers and regulators to develop clear guidelines and regulations to ensure that the use of surveillance technology is transparent, and accountable, and respects individuals' rights and liberties. Moreover, individuals need to be aware of the potential risks and take steps to protect their privacy and security in an increasingly surveilled world.

\section{Focus on Ethical Frameworks}
In this Section, the proposed Ethical Frameworks, of core importance for Ethic Sciences, will be analyzed through their core principles. Consequentially, a comparison between their assumptions, axioms, and implications will be proposed. In Table \ref{tab:philoso}, a direct comparison of the key points of the considered philosophical theories is presented \cite{morality}.

Both the phenomenon described in previous sections, i.e. target advertising in Section \ref{ads} and military technology in Section \ref{war}, are analyzed through the ethical perspectives.
\begin{table}[ht]
    \centering
    \begin{tabular}{|p{38mm}|p{38mm}|}
        \multicolumn{2}{c}{Comparison between the ethical frameworks} \\
        \hline
        \textbf{Utilitarianism} & \textbf{Deontology} \\
        \hline 
        Jeremy Bentham & Immanuel Kant \\
        Other & Self \\
        Utilitarianism Principle & Categorical Imperative \\
        Consequentialism & Motivationalism \\
        Utility Mathematical Calculation & Logical Inference \\
        Actions judged only on the basis of their consequences & Actions univocally judged as right or wrong for intrinsic nature and attributes \\
        Ethical behavior produces the greatest good for the greatest number & Ethical behaviour is identified by good will behind it \\
        \hline
    \end{tabular}
    \caption{Comparison between Utilitarian and Deontological frameworks. The most representative philosophers of such theories are respectively Jeremy Bentham and Immanuel Kant. Utilitarianism focuses on the welfare of the majority, meaning "Other", judging actions exclusively on a consequence-based approach. On the opposite side, Kant focuses on the "Self". Actions are judged on the basis of axioms, their inner attributes and logical inference, under the assumption of good will of the individual.}
    \label{tab:philoso}
\end{table}

\begin{figure}[t]
\centering
\includegraphics[trim=0mm 0mm 0mm 0mm,clip,width=\linewidth]{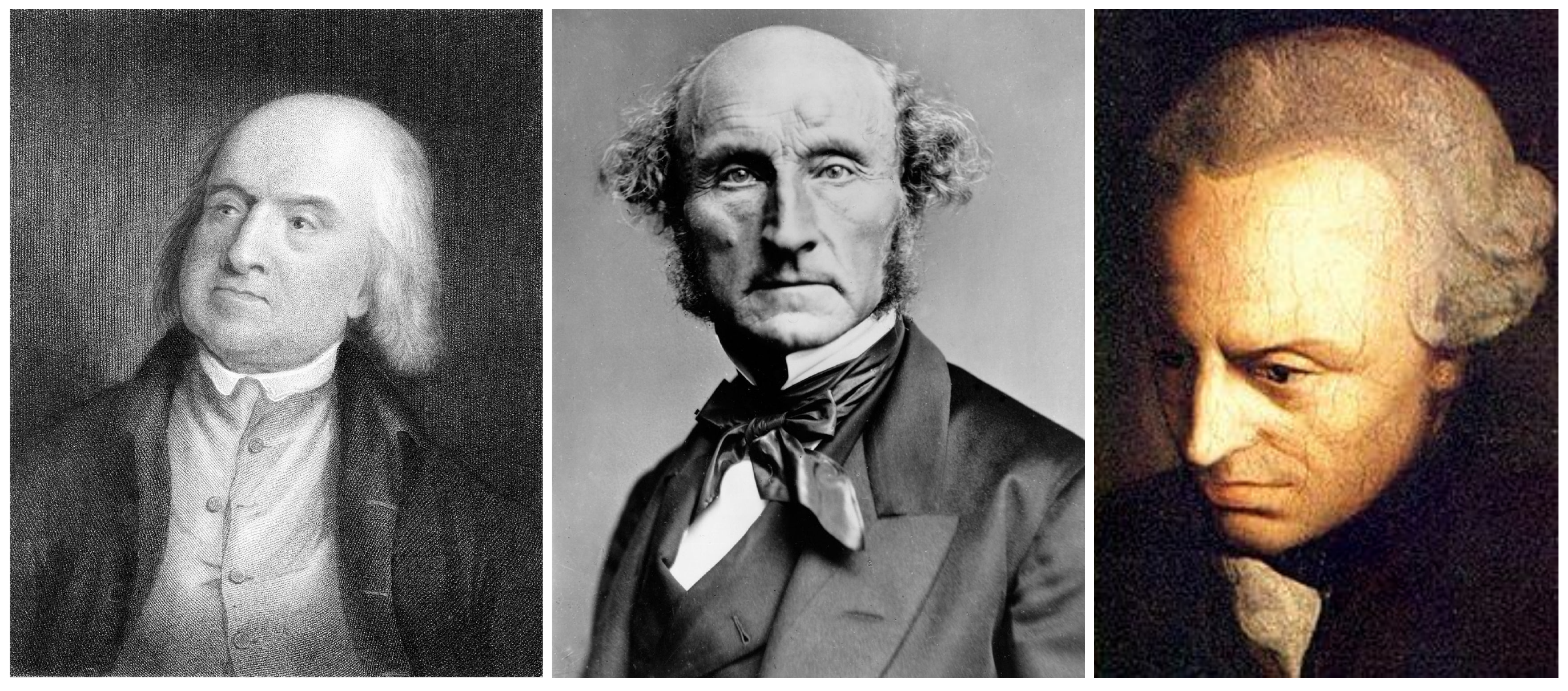}
\caption{From left to right. Jeremy Bentham, the founder of Utilitarianism, whose core idea of value theory was founded on hedonism, meaning that only pleasure is in itself the essence of good and towards which all other things represent simply an instrument. John Stuart Mill, a second representative of Utilitarianism, whose core idea was the Freedom principle, implying that the attainment of pleasure is pure and good as long as it does not affect the pleasure and well-being of others. Immanuel Kant, one of the most influential and important philosophers in history, founded its philosophy on criticism and moral laws derived from the categorical imperative.}
\label{phil}
\end{figure}
\subsection{Utilitarian Framework}
\label{util}
In this Section, an analysis under the light of Utilitarian principles is performed. Jeremy Bentham is best known for his theory of utilitarianism, which argues that the ethical value of an action is determined by its ability to promote the greatest happiness for the greatest number of people. Bentham believed that pleasure and pain were the ultimate factors that motivated human behavior, and that rational decision-making should always aim to maximize pleasure and minimize pain. In the modern era, Bentham's utilitarian ethic has been applied to various aspects of society, including business practices. One example of this is the rise of surveillance capitalism, which refers to the use of data collection and analysis to tailor advertising and other marketing strategies to individual consumers. John Stuart Mill is the second main proponent of utilitarianism, and also his ideas are particularly relevant, as he also emphasized the importance of individual freedom and autonomy, and argued that the state should only interfere with individual liberty in order to prevent harm to others.

From a utilitarian perspective, surveillance capitalism could be seen as a way to maximize happiness by providing consumers with products and services that are tailored to their individual needs and desires. However, there are also concerns that this practice can be exploitative and invasive, particularly if users are not aware of the extent to which their data is being collected and used. Bentham himself was a proponent of the use of surveillance as a means of promoting social welfare. He believed that constant monitoring and evaluation of human behavior could be used to identify and address social problems, such as crime and poverty. However, Bentham also recognized the potential for abuses of power in a system of surveillance. He argued that such a system should be subject to strict oversight and transparency in order to prevent its use for nefarious purposes. In the context of surveillance capitalism, Bentham's emphasis on transparency and oversight could be applied to ensure that users are fully aware of how their data is being collected and used, and that companies are held accountable for any misuse of that data. According to Mill, in the case of surveillance capitalism, the collection and use of personal data raise significant concerns about individual privacy and autonomy. Mill would likely argue that individuals have a right to control their personal data and that companies and other organizations should not be able to collect and use that data without explicit and informed consent. Moreover, Mill was concerned about the potential for harm to individuals and society that could arise from unchecked power, whether that power is held by the state or by corporations. In the case of surveillance capitalism, there is a risk that the power and influence of corporations could become too great, leading to the exploitation of individuals and the erosion of democracy.

According to the classical Utilitarian framework, propaganda, target advertisements, entertainment, and in general all the new technological goods provided by the Surveillance Capitalist society can be seen as a source of pleasure for the majority of people that makes use of it. This type of business allows to better fit the tastes and preferences of users. Therefore, from a superficial point of view, we can say that it helps in providing happiness for the majority. Besides, from the point of view of business companies, whose only goal is to make successful business, this phenomenon does not produce any damage or hurt a large number of people as data are merely used to cluster consumers and realize a more efficient business \cite{util1}. Furthermore, another important element that must be underlined is that the targeted promotions and offers are typically in agreement with the user preferences, so often they are not perceived as annoying. 

The core challenges are how to quantify the pleasure that is provided and how to weigh the portion of people that eventually derives pleasure, considering the optic “greatest pleasure for the most”. From another perspective, i.e. one of the users (and consumers), the harm can be perceived as a violation of privacy and most of all as a violation of the user’s free will and intention. The first unfairness was that the typical user was not aware, at the beginning, of this process, of the fact that clicking on a specific ad, then it would have been classified in a targeted group of people with a shared interest. Once awareness about the fact that explicitly declared preferences has been obtained, then a second unfairness raises. This is the fact that users and the vast majority of people are exploited and abused because they don’t share only what they think they deliberately put on social networks for example. Much other information that they don’t want to share without their consent is stolen and this is the expression of a hidden and sneaky form of abuse \cite{util3, zub}. Consequently, social media lead to an excessive liberalization of the public sphere and its access to private space and the other way around. Sharing online experiences, emotions, and important life events are free decisions that people make. A personal emotion or an event that is shared online by the user implies that almost always its experience becomes associated with a value, in the sense that is a quantifiable money resource for companies \cite{util2}. 

In the context of military technology, surveillance and monitoring can be used to gather intelligence and prevent threats to national security. However, there are concerns about the use of such technologies to violate individual rights and liberties. From a utilitarian perspective, surveillance and monitoring can be justified if they contribute to the overall happiness and welfare of society. However, Bentham recognized the potential for abuses of power in such systems and argued for strict oversight and transparency to prevent their misuse. Bentham's emphasis on transparency and oversight could be applied to ensure that individuals are fully aware of how their data is being collected and used and that companies are held accountable for any misuse of that data. Additionally, utilitarian ethics would require consideration of the potential negative consequences of such systems on individuals and society, such as the erosion of privacy and autonomy. In the context of military technology and politics, Bentham's utilitarian ethic would require consideration of the potential benefits and harms of surveillance and monitoring for national security. While surveillance and monitoring may be necessary for preventing threats and protecting citizens, utilitarian ethics would require careful consideration of the potential negative consequences on individual rights and liberties \cite{util3, zub}.

In conclusion, Jeremy Bentham's utilitarian ethic provides a framework for understanding the ethical implications of surveillance capitalism and targeted advertising. While this practice has the potential to maximize happiness by tailoring products and services to individual preferences, it also raises concerns about privacy and exploitation. Bentham's emphasis on transparency and oversight can help to address these concerns and ensure that surveillance capitalism is used in a way that maximizes the greatest happiness for the greatest number of people.
\subsection{Deontology Framework}
\label{duty}
In this Section, the analysis is performed under the light of Duty ethics principles. In the context of surveillance capitalism and targeted advertising, Kant's ethical theories have important implications for individual privacy and autonomy. According to the classical Kantian framework, in particular, considering the second categorical imperative, the reciprocity principle, we can say that the lack of autonomy that is induced by Surveillance Capitalism is completely unjustified and morally wrong. Here, the core point is that there should be respect for people’s moral autonomy in making their own choices, and manipulation is totally against this principle. Furthermore, trying to manipulate people’s ideas and decision-making process is disrespectful to the moral autonomy of reasoning, treating them as pure means \cite{duty 3}. This approach brings society to a deep context of social inequality as the majority of people is a victims of this unfairness and commodification. From a Kantian perspective, the collection and use of personal data without explicit and informed consent violate the inherent dignity of individuals, as it treats them as mere means to an end. For Kant, ethical actions must be grounded in respect for individual autonomy and dignity, and the use of personal data for targeted advertising undermines this respect. This is because it treats individuals as mere objects to be manipulated for commercial gain, rather than as autonomous and rational beings capable of making their own choices.

First of all, at the beginning people were unaware of this intrinsic mechanism. Nowadays, the perception of this phenomenon is more widely spread, and the awareness of the real implications is not completely known by the majority \cite{duty4}. Besides the intention of manipulation, also the lack of information of people goes in contrast with the reciprocity principle and the right of moral autonomy of people. On the other side, from the opposite point of view of analysis, there is the justification and original reason for which Surveillance Capitalism was born and developed in this direction, talking about social order and State Security \cite{duty1}. Under this perspective, we could consider that the universal law could be read as “Allowing people to secretly participate in criminal organizations performing attacks of an economic or terrorist or military nature is acceptable”. If this were to be a universal law, then people would be afraid of every aspect of society and this would lead to anarchy. No government would sustain a policy like this, because it would never be voted. This kind of maxim cannot be universalized, so it is clear that a form of control is always needed, desirable, and accepted by people, to keep and support social order and welfare. The issue in this delicate topic is borderline, meaning that, if we accept that control is allowed, then who deserves to be controlled, and what are the criteria for which a person can be selected to be controlled remain only some of the open questions of society. 

In this scenario, the Kantian framework would lead into contradiction, since, from one side it would justify the need for control because universally speaking it is true that attacks of any type that hurt people must be prevented, on the other side it would not accept the lack of autonomy of people, the manipulation and disrespect of moral reasoning. In practice, the fault in this line of reasoning is that strict and rigid adherence to moral rules can bring blindness and disruptive effects \cite{duty2}. The key idea is that it is essential to inform people, and this is connected to the importance of human rights. The action of controlling, perhaps, is not morally incorrect in itself, but in the moment in which people that are the object of control are not aware of it. 

Moreover, Kant was concerned with the principle of universalizability, which holds that ethical principles should be applicable to all individuals in all circumstances. In the case of surveillance capitalism and targeted advertising, this means that companies should not collect or use personal data in ways that they would not want to be treated themselves. Kant argued that ethical actions must be guided by moral duties, such as the duty to respect the inherent worth and dignity of every individual. In response to these concerns, Kant would likely argue for the development of laws and regulations that protect individual privacy and autonomy, while still allowing for the benefits of data-driven advertising and marketing. He would also emphasize the importance of informed and explicit consent in the collection and use of personal data, as well as the principle of transparency in corporate practices.
\subsection{Direct comparison between Utilitarianism and Deontology}
\label{comp}
Considering the first scenario, described in Section \ref{ads}, the first and immediate difference that can be underlined is that while Utilitarianism focuses mostly on pleasure and happiness, and could justify the means of Surveillance Capitalism to increase pleasure for the most, the Kantian framework will never justify such a politics with that idea. Kantian theory would underline the subjectivity of pleasure and happiness for each individual, and the difficulty in objectively measuring the effects (that were themselves the issues raised in Section \ref{ads}). Kant would not justify the abuse and violation of privacy for the business purposes of companies merely through the idea of making money without harming people. According to the pure sense of duty, such use of Surveillance Capitalism, keeping people unaware, would be morally incorrect, and people should be able to understand it. Furthermore, treating human beings as means only for personal and selfish interests without informing is a disrespect of human dignity according to Duty ethics. It must be noticed that, according to the Kantian framework, there is no violation of the reciprocity principle in the simple act of making a targeted advertisement because the individual is not prevented by the possibility of determining what is morally right through reasoning. The pure Kantian framework would agree with a society in which users' implicit data are retrieved and used with their awareness, if and only if the final goal is to make business. In this case, there is the economic scope of industry without leading humans' moral right to reason and their autonomy; this is reached by making them aware and using these data with the exclusive goal of selling more products and increasing sales. Furthermore, and most important, Kant would agree if there was the certainty of keeping data anonymous, as most research works are studied nowadays. On the contrary, Utilitarian framework will be more focused on the social welfare meaning that the greatest happiness must be produced for the majority, and will agree in assuming that this happiness is derived from the possibility of each individual to better satisfying their own tastes and preferences, by discovering new places, items or possibilities that are in line with its desires. Once the awareness about the fact that companies stored data, even implicit data retrieved by the user activity, has been gained, then the idea of abuse is not more applicable to this scenario and therefore, this behavior is acceptable because it does not harm anyone, especially if the target advertisement would get more and more flexible, meaning that the user has the possibility to explicitly to deny such a use of personal data to realize target advertisement. Indeed, Bentham's utilitarian ethical theory would evaluate the use of target advertising in surveillance capitalism based on its ability to promote overall happiness or utility. If targeted advertising leads to a net increase in happiness or utility by providing people with more relevant and useful information, it might be considered morally justifiable from a utilitarian perspective. However, the downside of this approach is that it could lead to privacy violations and a loss of individual autonomy, which could outweigh the potential benefits. Kant's duty ethic, on the other hand, would likely argue that the use of target advertising is inherently immoral because it fails to treat individuals as ends in themselves and instead treats them as mere means to an end. According to Kant, individuals have a right to privacy and should be treated as moral agents capable of making their own decisions. Therefore, any action that treats individuals as mere means to an end, such as using their personal data to manipulate their behavior through targeted advertising, would be considered morally impermissible.

Considering the second scenario, described in Section \ref{war}, the main differences can be immediately underlined by saying that if the scope is the one of preserving people from attacks of any nature, then according to the Utilitarian framework any action would be justified because the end justifies the means and the Surveillance has no a direct impact (meaning harming) on people. It is more of an invisible control that is needed for the welfare of the most, so it must be accepted. This idea is not true anymore under the Kantian perspective, in which, as described in section \ref{duty}, the Universal principle according to which such behavior, would be considered acceptable leads to a contradiction. The key point and solution to this contradiction, to make the framework effective would be again the solution of lack of awareness of people. The idea of defining a threshold, or a metric, that can state if a human being deserves or not to be controlled is arbitrary, questionable, and mistakable. It would lead to errors and subjectivity, so it is helpful, but will not be a solution to the contradiction raised in the framework. On the contrary, by informing people, the aspects of manipulation and lack of autonomy fade, so the action of controlling, if Surveillance Capitalism was not devoted to influencing preferences and avoiding autonomy, but simply to the overall welfare, would be morally correct. It must be noticed, that an important difference while analyzing the effect of Surveillance Capitalism in society is the system of the considered country. If totalitarian regimes see these instruments as a way to enforce their control on the population and this is not morally acceptable under any perspective, both Kantian and Utilitarian, democratic and capitalist systems have valued to recur to Surveillance for reasons more connected to its rationality and efficiency, if well applied (normalized in the respect of people). Indeed, Bentham's utilitarianism might support the use of military technology and politics to collect data and monitor people's behavior if it leads to a net increase in happiness or utility. For example, if the government's use of surveillance technology can prevent crime and protect citizens' safety, it might be considered morally justifiable from a utilitarian perspective. However, the downside of this approach is that it could lead to privacy violations and a loss of individual autonomy, which could outweigh the potential benefits. On the other hand, Kant's duty ethic would likely argue that the use of surveillance technology is inherently immoral because it violates people's autonomy and fails to treat them as ends in themselves. According to Kant, individuals have a right to privacy and should be treated as moral agents capable of making their own decisions. Therefore, any action that treats individuals as mere means to an end, such as collecting data without their consent, would be considered morally impermissible.

\section{Conclusions}
\label{end}
This analysis has brought important conclusions. In both frameworks, we have seen how important the aspect of awareness is, even if to front different issues, according to the specific ethical framework. Another important point that was raised, considering the first scenario, is that an apparently less damaging and “innocent”, meaning harmless, action, like the one of the targeted advertisement, can be as well perceived as very much annoying and is not at all less serious than more evident forms of control. It can be seen as a pleasant form of intrusion in the Utilitarian framework, perhaps, or as controversial in the Kantian framework, but in both cases needs to be controlled, normalized, not abused. 

It is very important that research today is going in the direction of making data anonymous in order to allow business analysts to take profit from it because it means that people are sensitive to topics of privacy, autonomy and surveillance and that one-day commercial benefits and welfare, increasing business, could be derived in the total respect of the human being. The most important thing is to make IT able to protect rights. Considering the second scenario, the last point that was raised was the importance of social and national Security, together with the tendency of manipulation that can derive from the owning of important means like the ones of IT. In this case, social education would prevent the improper use of the control that can be applied and brought the Kantian framework into contradiction, always considering that inappropriate behaviors are always possible and the goal is the one of minimizing unfairness.

While the issues of privacy, autonomy, and surveillance have been around for a long time, they have become more pressing in recent years due to the increasing amount of personal data that is being collected and analyzed by companies and governments. As technology continues to advance, the potential for abuse of this data becomes even greater, and it is essential that we take steps to protect individual rights and freedoms. One possible solution is to increase transparency and accountability in data collection and usage. This can be achieved through measures such as data anonymization, which allows analysts to work with aggregated data without compromising the privacy of individual users. Additionally, regulations such as the General Data Protection Regulation (GDPR) in Europe and the California Consumer Privacy Act (CCPA) in the United States aim to give individuals greater control over their personal data and require companies to be more transparent about their data practices. Another important consideration is the role of education in promoting responsible data usage. This includes not only educating the general public about their rights and how to protect their privacy but also providing training to professionals in fields such as data science, who need to be aware of the ethical implications of their work. By promoting a culture of responsible data usage and creating incentives for companies and governments to prioritize individual rights, we can work towards a future where technology is used to benefit society as a whole, rather than just a few powerful actors.

In conclusion, the issues raised in this essay highlight the complex ethical considerations that arise in the context of data usage and control. While there is no easy solution to these challenges, it is clear that we need to prioritize individual rights and work towards greater transparency, accountability, and education in order to minimize the potential for abuse and promote a more just and equitable society. The practice of surveillance capitalism raises important ethical questions that must be carefully considered in the context of rapidly evolving technology and changing social norms. By examining the phenomenon from multiple ethical frameworks and real-world examples, researchers and practitioners can gain a more comprehensive understanding of the complex trade-offs involved.

\begin{acknowledgment}
I want to thank professor Viola Schiaffonati for supporting the development of this work, giving food for thoughts and raising awareness for many ethical issues related to the IT profession. 

Computer ethics also allows us to consider the diverse perspectives and values of various stakeholders, including users, communities, and society as a whole. This helps to ensure that our research and practice align with broader ethical and social considerations, and can contribute to positive social and environmental outcomes. Furthermore, it promotes the development of a culture of ethical responsibility and professionalism in the technology sector. By emphasizing the importance of ethical decision-making and accountability, we can help to create a more just and equitable society where technology serves the common good.

I think this consciousness is essential to improve us as engineers.
\end{acknowledgment}

%
% ...




{\footnotesize
\begin{thebibliography}{0}
  \bibitem{Surveillance1} {JB Foster, RW McChesnet. Monthly Review 2014. Surveillance capitalism: Monopoly-finance capital, the military-industrial complex, and the digital age}
  \bibitem{Surveillance2} {JE Cohen. Cambridge Handbook of Surveillance Law, eds. David. 2017. Surveillance vs. Privacy: Effects and Implications} {\scriptsize \newline$[$\url{https://papers.ssrn.com}$]$}
  \bibitem{Surveillance3} {SM West. Business \& society, 2019. Data capitalism: Redefining the logics of surveillance and privacy} {\scriptsize \newline$[$\url{https://journals.sagepub.com}$]$}
  \bibitem{ads1} {O Barbu. Procedia-Social and Behavioral Sciences, 2014. Elsevier. Advertising, microtargeting and social media}
  \bibitem{ads2} {CE Tucker. 2013, Social Networks, personalized advertising, and privacy controls} 
  \bibitem{ads3} {RG Pensa, G Di Blasi. Expert Systems with Applications, 2017. Elsevier. A privacy self-assessment framework for online social networks}
  \bibitem{ads4} {J Raudeliūnienė, V Davidavičienė, M Tvaronavičienė. Sustainability, 2018. Evaluation of advertising campaigns on social media networks} {\scriptsize \newline$[$\url{https://mdpi.com}$]$}
  \bibitem{war2} {D Muench, B Hilsenbeck, H Kieritz. Optics and Photonics for Counterterrorism, Crime Fighting, and Defence XII. 2016. Detection of infrastructure manipulation with knowledge-based video surveillance} {\scriptsize \newline$[$\url{https://spiedigitallibrary.org}$]$}
  \bibitem{war3} {J Burkell, PM Regan. Internet Policy Review, 2019. Voter preferences, voter manipulation, voter analytics: policy options for less surveillance and more autonomy} {\scriptsize \newline$[$\url{https://econstor.eu}$]$}
  \bibitem{war4} {WJ Jie, 2020. Between Surveillance and Security: The Protection from Online Falsehoods and Manipulation Bill (POFMA)} {\scriptsize \newline$[$\url{https://scholarbank.nus.edu.sg}$]$} 
  \bibitem{util1} {C Ess, C Kwok, NK Chan, M Bay. AoIR Selected Papers of Internet Research, 2020. LEGAL AND ETHICAL PERSPECTIVES ON (BIG) DATA, PLATFORMS, AI AND ALGORITHMS} {\scriptsize \newline$[$\url{https://journals.uic.edu}$]$}
  \bibitem{util2} {K Plangger, M Montecchi. Journal of Interactive Marketing, 2020. Elsevier. Thinking Beyond Privacy Calculus: Investigating Reactions to Customer Surveillance}
  \bibitem{util3} {M Andrejevic. Surveillance \& Society, 2019. Automating surveillance} {\scriptsize \newline$[$\url{https://ojs.library.queensu.ca}$]$}
  \bibitem{duty1} {Shoshana Zuboff, 2015. Big other: surveillance capitalism and the prospects of an information civilization. Journal of Information Technology}
  \bibitem{duty2} {JM Jacobs, QB Fox. 2019. Surveillance capitalism and its (un) intended consequences} {\scriptsize \newline$[$\url{https://ses.library.usyd.edu.au}$]$}
  \bibitem{duty3} {N Couldry. Journal of Information Technology \& Politics, 2017. Taylor \& Francis. Surveillance-democracy}
  \bibitem{duty4} {S Sevignani. Historical Social Research/Historische Sozialforschung, 2017. JSTOR. Surveillance, classification, and social inequality in informational capitalism: The relevance of exploitation in the context of markets in information}
  \bibitem{morality}{Yama, Hiroshi. (2020). Morality and Contemporary Civilization. 10.4018/978-1-7998-1811-3.ch004.}
  \bibitem{zub} {Shoshana Zuboff. The age of surveillance capitalism. The fight for a human future at the new frontier of power. VPRO Documentary. 2019} {\scriptsize \newline$[$\url{https://www.youtube.com/watch?v=hIXhnWUmMvw}$]$}
\end{thebibliography}}
\end{document}